\documentclass[prx,showkeys,superscriptaddress,preprint]{revtex4-1}   
\usepackage{amssymb}
\usepackage{amsmath}
\usepackage{graphicx}
\usepackage[english]{babel}
\usepackage{subfig}
\usepackage{color}
\usepackage[usenames,dvipsnames]{xcolor}
\usepackage{hyperref}
\newcommand{\cte}{\color{Black}}
\newcommand{\be}{ $J_0$}
\newcommand{\etal}{\emph{et al. }}
\newcommand{\vgr}{\emph{v.gr. }}
\newcommand{\eg}{\emph{e.g. }}
\newcommand{\ie}{\emph{i.e. }}
\newcommand{\bgj}{Bessel-Gauss }
\newcommand{\kphotonics}{Photonics Laboratory, King Abdullah University of Science and Technology \emph{(KAUST)}, Thuwal 21534, Saudi Arabia}
\newcommand{\kfupm}{King Fahd University of Petroleum and Minerals \emph{(KFUPM)}, Dhahran 31261, Saudi Arabia}

\begin{document}
\title{Generation of \be-\bgj Beam by an Heterogeneous Refractive Index Map}
\author{\firstname{Dami\'an P.} \surname{San-Rom\'an-Alerigi}}
\email[email: ]{damian.sanroman@kaust.edu.sa}
\affiliation{\kphotonics}
\author{\firstname{Tien K.} \surname{Ng}}
\affiliation{\kphotonics}
\author{\firstname{Ahmed} \surname{Benslimane}}
\affiliation{\kphotonics}
\author{\firstname{Yaping} \surname{Zhang}}
\affiliation{\kphotonics}
\author{\firstname{Mohammad} \surname{Alsunaidi}}
\affiliation{\kphotonics}
\affiliation{\kfupm}
\author{\firstname{Boon S.} \surname{Ooi}}
\affiliation{\kphotonics}

\begin{abstract}

{\cte In this paper, we present the theoretical studies of a refractive index map to implement a Gauss to \be-\bgj convertor. We theoretically demonstrate the viability of a device that could be fabricated on  $Si/Si_{1−y} O_y /Si_{1−x−y}Ge_x C_y$ platform or  by photorefractive media. The proposed device is $200\mu m$ in length and $25\mu m$ in width, and its refractive index varies in controllable steps across the light propagation and transversal directions. The computed conversion efficiency and loss are $90\%$, and $-0.457dB$, respectively. The theoretical results, obtained from the beam conversion efficiency, self-regeneration, and propagation through an opaque obstruction; demonstrate that a 2D graded index map of the refractive index can be used to transform a Gauss beam into a {\be}-\bgj beam.} To the best of our knowledge, this is the first demonstration of such beam transformation by means of a 2D index-mapping which is fully integrable in silicon photonics based planar lightwave circuits (\emph{PLC}). The concept device is significant for the eventual development of a new array of technologies, such as micro optical tweezers, optical traps, beam reshaping and non-linear beam diode lasers.
\newline 

\noindent \bf{Manuscript  published at the \emph{Journal of the Optical Society of America A},  \href{http://dx.doi.org/10.1364/JOSAA.29.001252}{JOSA A, Vol. 29, Issue 7, pp. 1252-1258 (2012)}}

\end{abstract}
\keywords{Bessel beam, transformation optics, electromagnetic inverse problem, inverse pde problem}
\maketitle

\section{Introduction}

{\cte In 1987 Durnin introduced non-diffracting $J_p$ Bessel  beams as  members of a family of solutions to the homogeneous Helmholtz equation, with distinctive properties, \eg their transverse profile does not change as the beam propagates in free space and exhibit self-regeneration when disturbed by non-transparent obstacles \cite{1}.} Among them the zero order, $J_0$,  Bessel beam has captured most attention due to its peculiar intensity distribution, which is focused on the propagation axis, besides non-diffracting propagation \cite{2}.\\

{\cte Whilst mathematically viable, Bessel beams are not square integrable, \ie the beam contains infinite energy, thus, rendering the ideal solution infeasible \cite{6,26}. Yet good approximation exists, namely \bgj beams. Introduced by Gorin \etal  \cite{26} these solutions to Helmholtz homogeneous equation resemble the ideal $J_p$ Bessel beam with the sole difference that they bear finite power; and, therefore,  can be realized experimentally, while retaining the primordial characteristics of self-reconstruction and diffraction-free propagation for lengths of interest to many optical applications, \eg optical tweezers and microscopy \cite{26,3,4,5}.}\\

Experimentally, Durnin \etal \cite{7} were the first to show that it is possible to generate \bgj beams by means of a circular slit and a lens placed one focal length away. Hakola \etal \cite{8} proposed a Nd:YAG cavity with a diffractive mirror to transform the Gauss beam from a pump laser into a \bgj beam of arbitrary order; Arriz\'on \etal \cite{9} and Otero \cite{10} independently used holograms to alter Gauss beams and produce zero and first order \bgj beams; lately, Zhan \cite{11} has shown that it is possible to generate them by using a radially polarized beam and surface plasmon resonance. Alternatively  Cong \cite{2} used phase elements to generate zero order \bgj beams. {\cte Recently novel designs have used waveguides to produced diffraction-free beams, \vgr Canning \cite{27} presented a Fresnel waveguide as a diffraction-free mode generator, and  Ilchenko \etal \cite{29} experimentally showed that cylindrical waveguides  can produce truncated \bgj beams;  and Tsai \etal \cite{30} showed that is possible to use an acoustic tunable lens to generate this family of beams.} \\

The afore mentioned methods produces high quality beams. Nonetheless their design and fabrication on  photonic integrated circuits (PIC) or planar lightwave circuits (PLC)  would involve complex processing stpdf, rendering the integration into most PICs platforms impractical. {\cte It is a matter of debate whether  resonator/waveguide designs (see for example \cite{27,29}) could be used for this purpose. In our experience on such devices, the quality of the beam, and conversion efficiency  critically depend on the coupling efficiency. Moreover the converted beam free space propagation is limited to $\sim 1mm$ \cite{29}.}  Hence the motivation to attain  a fully integrated Gauss to \be-\bgj micro-convertor based on current $Si$ foundry technology is justified. If viable, it could be readily integrated into semiconductor-chip-based optical elements of great importance to a manifold of optical applications. For instance they have been shown to increase the scanning resolution and tissue penetration depth of optical coherence tomography (OCT) systems by $\sim 50\%$ \cite{12}.\\

\section{Problem description and modeling}

{\cte In Durnin's primordial design, an impinging Gauss beam diffracts from the surface of an axicon lens \cite{4,7,13}, forming a converging wavefront and thus giving raise to the \bgj beam. Analytically it can be described by means of diffraction at the Fresnel limit. However accurate this method could be,  it presumes the outbound field to extent infinitely in the transversal direction in order to comply with the boundary conditions \cite{4}.  To avoid this complication, we will assume hereafter the output beam to be \be-\bgj as defined by Gori \etal \cite{26}, \ie the superposition of a Gauss and $J_0$-Bessel profiles.\\


It could be contended that only a PIC-integrable-axicon lens would suffice to achieve the desired conversion. Such lens, though, is equivalent to the fabrication of a continuous alteration of the index of refraction, albeit homogeneous, over a conic volume, posing practical and non-trivial integration problems when it is manufactured using the well-established silicon fabrication technologies, namely the difficulty to shape the conic surface. To alleviate this conundrum, we seek a practical profile of the refractive index that can act as feasible substitute for the axicon lens, suited for integration into current silicon foundry and manufacturing technologies.}\\

{\cte Before proceeding any further, it is important to remark that non-singular graded refractive indexes (GRIN) cannot yield perfect non-diffracting beams as proven by Hayata \cite{28}. However a refractive index mapping can potentially achieve high quality conversion from a Gauss beam to a \be-\bgj beam, which is different from the fundamental Bessel beam presented by Durnin \etal \cite{1}.}\\

\begin{figure}[b]
\centering
\includegraphics[width=.8\textwidth]{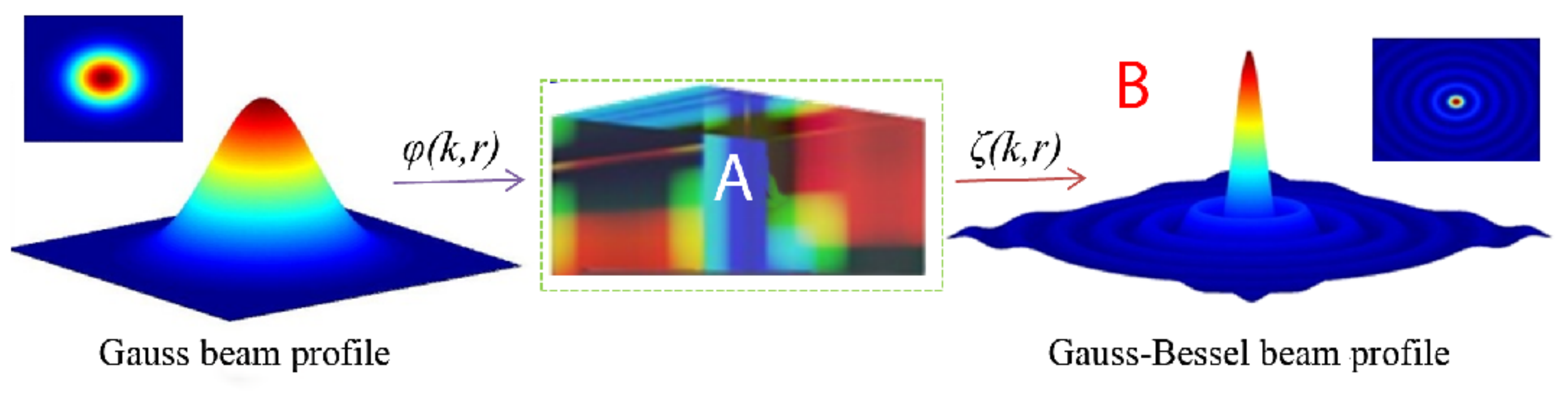}
\caption{Schematic representation of a Gauss to $J_0$-\bgj transformation \textit{via} a heterogeneous medium,  the transformation device in ragion $A$ has a refractive index map that needs to be calculated in order to echieve the desired beam transformation. The device is embedded in a homogeneous medium $B$, where Bessel and Gauss beams are known solutions to the Helmholtz equation. \label{fig.artistic}}.
\end{figure}

{\cte In Figure \ref{fig.artistic}, the input and output beams are known \emph{a priori}, and hence the problem is reduced to determining the heterogeneous refractive index in region $A$, which will result in the transformation of the input to output fields in region $B$.} This is the  optical inverse scattering problem \cite{14,15,16,25}.\\

Analytically, the input and output fields are described by harmonic functions of the form $\vec E_i=\vec \phi (\vec k,\vec r)e^{i\omega t}$ and $\vec E_o=\vec \zeta (\vec k,\vec r)e^{i\omega t}$, respectively; where  $\vec \phi$ and $\vec \zeta$  are solutions to the homogeneous Helmholtz equation. In the transformation region, medium $A$, the field is $\vec E_A=\vec \vartheta (\vec k,\vec r) \tau(t)$; where $\vec \vartheta$ is determined by the optical properties of the  medium.  If the material is  linear, then $\tau(t)=e^{i\omega t}$, and the problem narrows to determining the shape of  the permittivity, $\varepsilon$, and permeability, $\mu$, that realize the transformation $\vec \phi \to \vec \zeta$ after some finite propagation distance, $d$.\\

{\cte In reality, the manipulation of permeability, $\mu$, is complex. On that account the transformation medium $A$ is set to be  non-magnetic, with some constant value for $\mu$. Consequently,  Maxwell's equation for the electric field $\vec E$ for this region is written as:}

\begin{equation}\label{eq1}
\centering
\nabla ^2\vec E-\mu \varepsilon(\vec r)\frac{\partial ^2\vec E}{\partial t^2}=-\nabla \left(\frac{1}{\varepsilon }\vec E\cdot \nabla\varepsilon\right),
\end{equation}

\noindent where $\varepsilon(\vec r)$, the permittivity, is a real value function and $\vec r=(u,v,w)$ describes a position vector in the transformation space. Outside the device the material is homogeneous, \textit{i.e.} $\varepsilon(\vec r)=constant$ for all $\vec r \in B$, where both Gaussian and Bessel functions are well known solutions to equation \ref{eq1}. Inside the device, the permittivity is limited to the range achievable in practical fabrication methods:

\begin{equation}\label{epsilonmaxmin}
\centering
\varepsilon_{min} \leq \varepsilon(\vec r) \leq \varepsilon_{max},\quad \forall \ \vec r \in A.
\end{equation}
\\
\indent Recalling that if $E_A=\vec \vartheta(\vec r) \tau(t)$, then equation \ref{eq1} can be reduced to the time independent Helmholtz equation:

\begin{equation}\label{eq2}
\centering
\nabla^2{\vec \vartheta}  (\vec r)-\omega^2 \mu \varepsilon (\vec r)\vec{\vartheta} (\vec r)=\nabla \left(\vec{\vartheta} (\vec r)\cdot \nabla {\ln  \varepsilon (\vec r)}\right).
\end{equation}
\\
\indent Observe that if the field $\vec \vartheta(\vec r)$ is given, then the only unknown in equation  \ref{eq2} would be $\varepsilon(\vec r)$, which is the permittivity map that we wish to determine. Henceforth let us examine $\vec{\vartheta} (\vec r)$.\\

Denote by $S_i$ and $S_o$, the boundary surfaces between region A, and the incident $B_i$ and exit $B_o$ surfaces, respectively. Then  $\vec{\vartheta} (\vec r)$ can be split into the three regions:

\begin{equation}\label{eq3}
\centering
\vec\vartheta (\vec r) =\left\{ \begin{array}{c}
\begin{array}{cl}
\vec \phi (\vec r) & \quad \forall \quad \vec r \in  S_i, \\ 
\vec \psi (\vec r) & \quad \forall \quad  \vec r \in A, \\ 
\vec \zeta (\vec r) & \quad \forall \quad \vec r \in S_o, \end{array}
\end{array}
\right.
\end{equation}

\noindent where $\vec \psi:A\to R^3$ is the transforming field. Since we wish to control $\vec \vartheta$, and thence decrease the complexity of the  problem, we define the shape of $\vec \psi$ across  region $A$. Accordingly, we introduce the real valued functions ${{f,g}:A^3 \to [0,1]}$ and ${\vec \gamma:A\to R^3}$, $C^2$ continuous and square integrable, \ie smooth functions. These functions are  invoked to describe the transforming beam as a superposition of the desired input and output fields, in addition to an arbitrary field  $\vec \gamma$ that can be thought to represent the residual beam. The field $\vec \psi (\vec r)$ can be written now as:

\begin{equation}\label{eq4}
\centering
\vec \psi (\vec r)=f(\vec r)\vec{\phi} (\vec r) + g(\vec r) \vec{\zeta} (\vec r) + \vec \gamma(\vec r),
\end{equation}

\noindent to comply with the boundary condition $f(S_i)=g(S_o)=1,\ f(S_o)=g(S_i)=0$ and $\vec \gamma(S_i)=0$. Then substituting equation \ref{eq4} into equation \ref{eq2} results in a second order partial differential equation:
\begin{equation}\label{eq5}
\overbrace{\nabla^2\left(f(\vec r)\vec{\phi} (\vec r) + g(\vec r) \vec{\zeta} (\vec r)\right)}^{\vec \varpi} -\overbrace{\omega^2 \mu \cdot\left(f(\vec r)\vec{\phi} (\vec r) + g(\vec r) \vec{\zeta} (\vec r)\right)}^{\omega^2\mu \vec \psi}\cdot \varepsilon(\vec r)  =  \nabla\left[\left(f(\vec r)\vec{\phi} (\vec r) + g(\vec r) \vec{\zeta} (\vec r)\right) \cdot \nabla {\ln {\varepsilon (\vec r)}}\right].
\end{equation}

\indent Recalling vector calculus identities, the right hand side of equation \ref{eq5} can be reduced to:
\begin{equation}\label{eq6}
\nabla\left[\left(f(\vec r)\vec{\phi} (\vec r) + g(\vec r) \vec{\zeta} (\vec r)\right) \cdot \nabla  {\ln {\varepsilon (\vec r)}}\right]=\left(\vec \psi \cdot \nabla\right)\nabla {\ln {\varepsilon (\vec r)}}+\nabla_\psi  \left(\nabla {\ln{\varepsilon (\vec r)}}\cdot \vec \psi\right).
\end{equation}
\\
\indent Observe that in equation \ref{eq5}, $\vec \varpi$ and $\vec \psi$ are vector functions with known values across the three domains previously described (see equation \ref{eq3}). Thus by means of equation \ref{eq6} we can rewrite equation \ref{eq5} as a differential equation that describes the permittivity, $\varepsilon$, as:

\begin{equation}\label{eqforeps}
\omega^2\mu \vec \psi \varepsilon(\vec r) + \left(\vec \psi \cdot \nabla\right)\nabla {\ln {\varepsilon (\vec r)}}+\nabla_\psi  \left(\nabla {\ln{\varepsilon (\vec r)}}\cdot \vec \psi\right) = \vec \varpi.
\end{equation}
\\
\indent This is the nonlinear equation for the permittivity that we ought to solve within the boundary conditions and  fabrication limitations defined earlier:  constant permeability, $\mu $, refractive index range  $1.5\le n(\vec r)\le 3.5$, where $n(\vec r)=\sqrt{\varepsilon(\vec r)\mu}$; and fabrication minimum feature size $d$.\\   

\begin{figure}[b]
\centering
\subfloat[Transformation transition]{\label{figure.stages} \includegraphics[width=.45\textwidth]{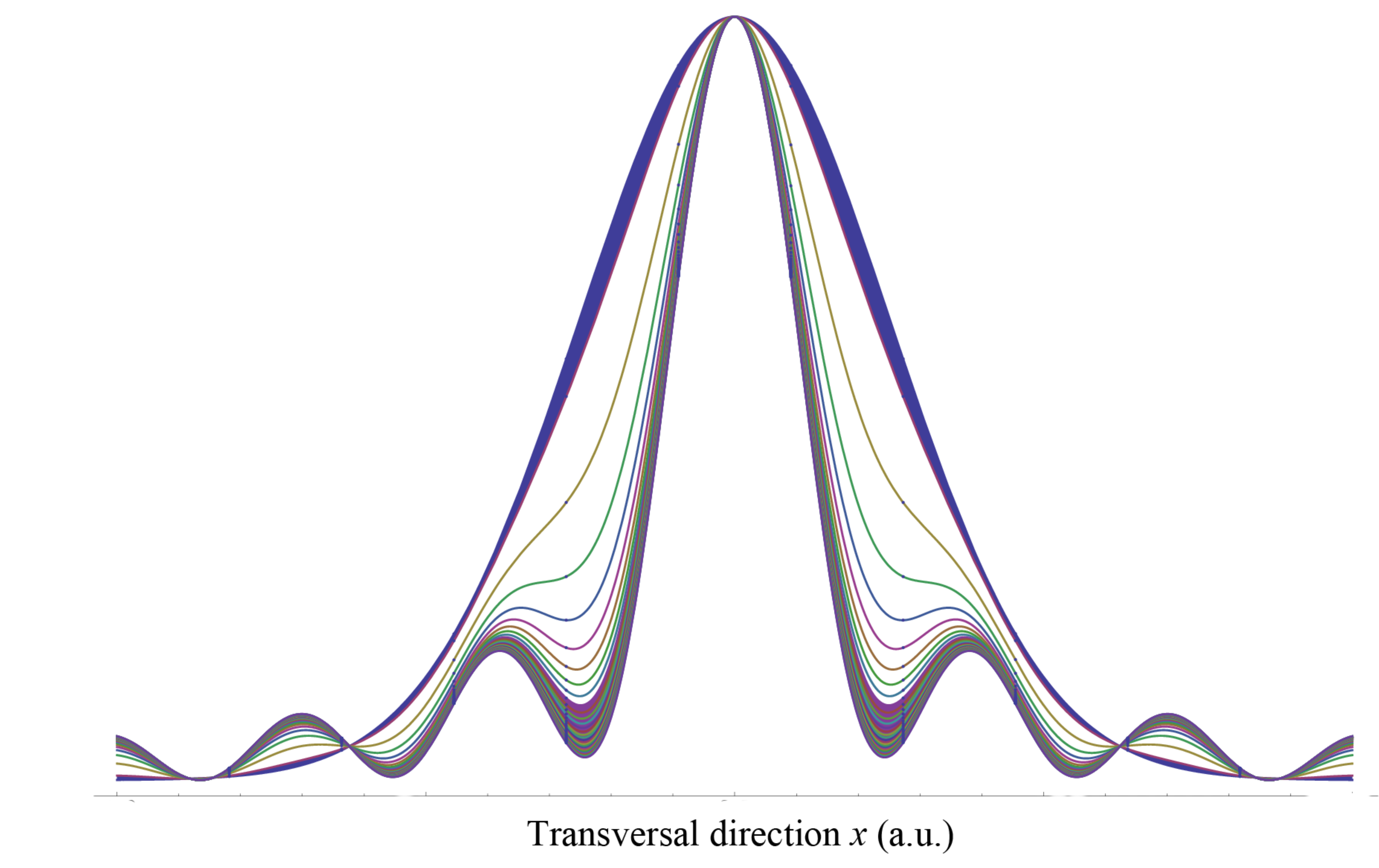}}
\qquad
\subfloat[Characteristic functions $f$ and $g$]{\label{fig.fg}\includegraphics[width=.45\textwidth]{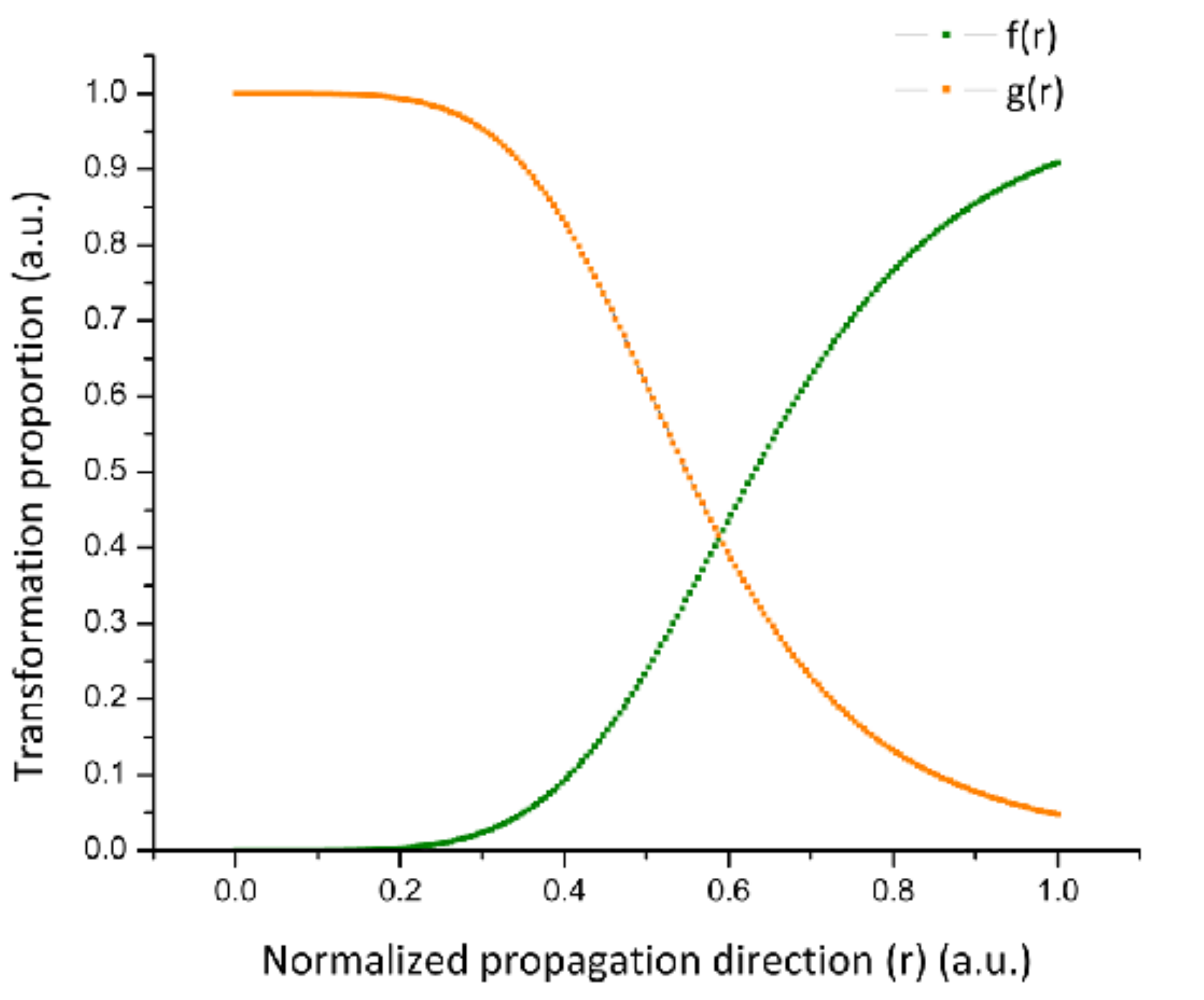}}
\caption{Illustration of (a) superposition of  transition stages in the transformation of  Gauss to \be-\bgj beams; and (b) the transformation functions $f$ and $g$.}
\label{fig.illustrationfg}
\end{figure}

{\cte The computation of an inverse problem as the one hitherto described by equation \ref{eqforeps} is not trivial, and analytical solutions are rarely found. Moreover, the problem described by the above equation is non-linear and ill-defined, \ie its solution is not unique and is dependent continuously on the data \cite{17,31}. As a result, many numerical schemes have been developed to elucidate the solution, \vgr the adjoint problem \cite{14,17}, linear sampling \cite{18,19,20}, variational methods \cite{21},  and lately transformation optics \cite{22,23,25}. To find a solution to this quandary, we developed a numerical technique based on the variational method and the adjoint problem.\\ 

We begin by noting that if the input and output beams are symmetrical around the propagation axis $z$, and does as well the field in the transformation region, then we can scale down the problem to two-dimensions: $z$ and $y$, the longitudinal and transversal direction, respectively. Concerning our matter at hand both Bessel and \be-\bgj beams are symmetric around the propagation axis; moreover the transformation field as described by the above equations is also symmetric around the axis $z$. Therefore, our problem can be scaled down to a two dimensional numerical scheme.\\

Accordingly, the simulation space is discretized on a grid size $d^z$ by $d^y$. The solver is given an initial guess of the permittivity map $\varepsilon_{ij}$ across the region $A$ at every point ${z_i,y_j}$. The solution to equation \ref{eq2} is computed via a variational approach as described in \cite{21}: at every point $z_i$ the transversal function of $\varepsilon_{ij}$ is determined by solving equation \ref{eqforeps} to the boundary conditions (see equation \ref{eq4}): $\vec \psi \left(y, z_{i−1}\right)$ and $\vec \psi \left(y, z_{i+1}\right)$; and the field at that point $\vec \psi \left(y, z_i\right)$. This iterative approach is carried for every  point $z_i$. Thus the transversal shape of $\varepsilon_{ij}$, and consequently that of $n_{ij}$, is determined to match the resulting field to the ideal transformation given by equation \ref{eq4}.\\

This method solely does not guarantee that the solution will be exempted of extreme or singular values, withal the variational method is especially sensitive numerical approximations \cite{21}. To account for the latter and work out the extrema in the refractive index, we implement a solution coming from the adjoint problem \cite{17}. The extreme and singular values are approximated to the closest value in the defined range. We then propagate the initial Gauss beam through the device and determine how much the field at every point diverges from the ideal transformation of equation \ref{eq4}, including as well the near and far fields. An iterative refinement on this grid regions follows, as described in \cite{14,17}, to match the numerical computation output to the ideal \be-\bgj beam. The beam propagation method is used to propagate the initial Gauss beam through the heterogeneous region $A$.\\

It is important to note that both, the variational and adjoint problem methods, while widely tested may render inexact solutions due to numerical approximations which can lead to instability and are in general related to the limited resolution of any numerical method. To counterweight this difficulties the general approach is to use smaller grid sizes even at the expense of computational time increasing exponentially.}\\

\section{Results.}

{\cte The transformation functions $f$ and $g$ are plotted in Figure \ref{fig.fg}. The initial guess for the permittivity map is a Gaussian profile in the transversal direction $y$ for all points along $z$, width $\sigma=5$ and a maximum peak of $\varepsilon_{max}=9$ or $n_{max}=3$. The initial beam in region $B_i$  follows a Gaussian function, with width $10 \mu m$ and normalized energy to unity}. \\

\begin{figure}[b]
\centering
\includegraphics[width=.7\textwidth]{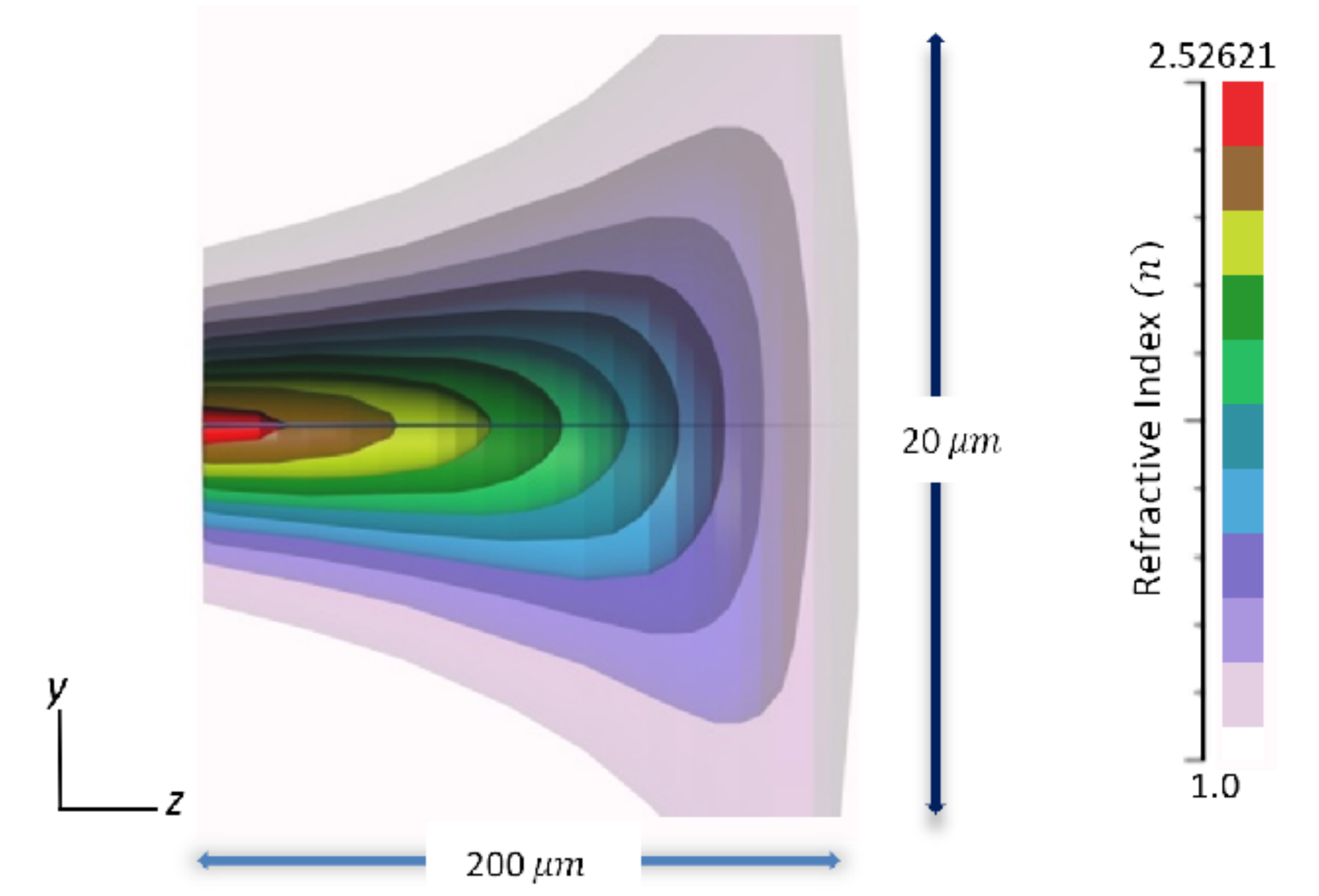}
\caption{Sketch of the proposed refractive index map for the conventor as defined by the transformation functions $f$ and $g$; with device length of $200 \mu m$ and width $25 \mu m$. Albeit here shown in $11$ steps, the refractive index varies smoothly between $1$ to $2.57$ in controllable steps. The device is rotationally symmetric around the propagation axis $z$.}
\label{fig.convertor}
\end{figure}

\begin{figure}[b]
\centering
\includegraphics[width=.7\textwidth]{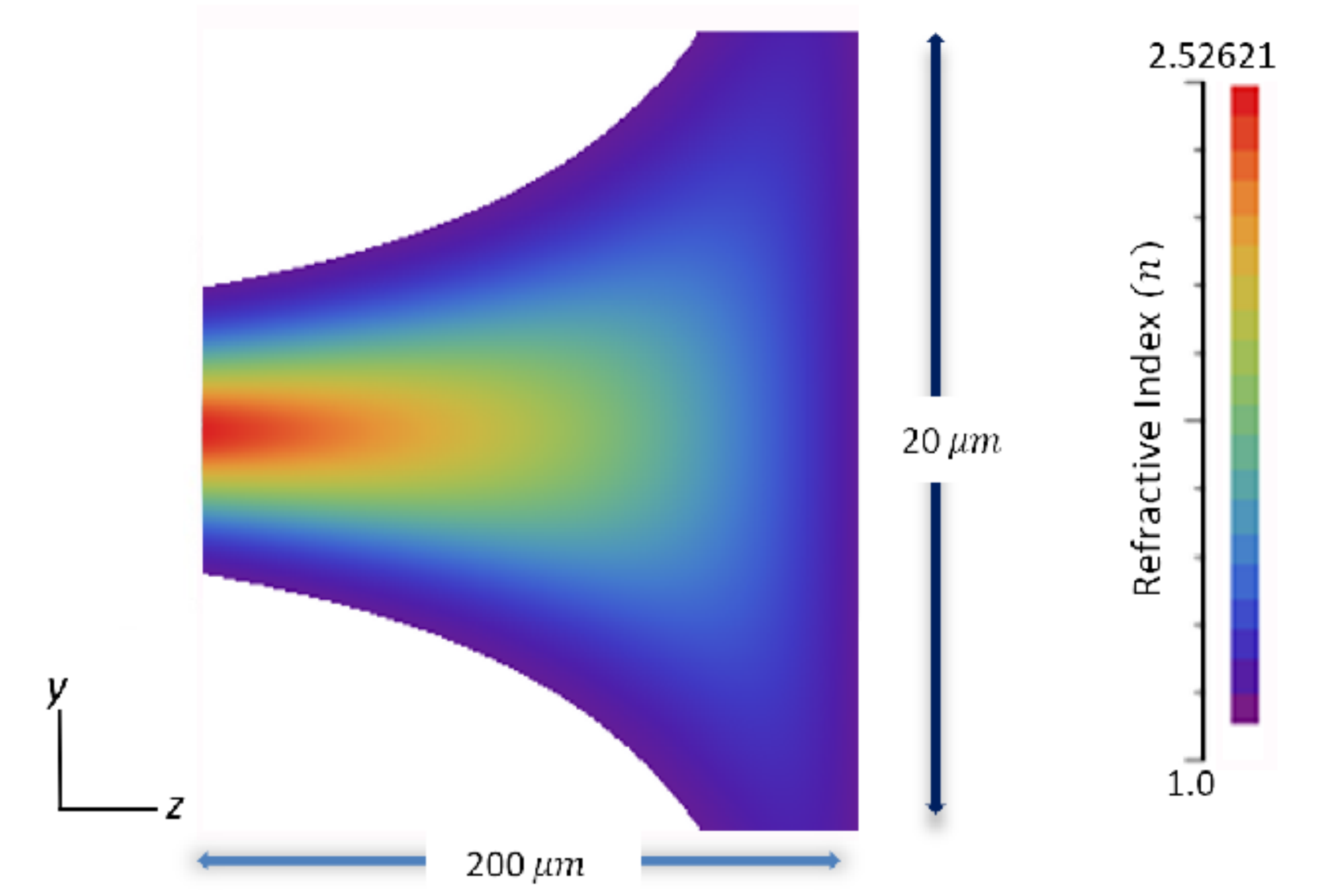}
\caption{Fine resolution refractive index map for the conventor, with minimum feature size $d=1nm$, as defined by the transformation functions $f$ and $g$. Device length is  $200 \mu m$ and width $25 \mu m$. The refractive index varies smoothly between $1$ to $2.57$ in $50$ controllable steps. A practical device could be fabricated on a $Si/Si_{1−y} O_y /Si_{1−x−y}Ge_x C_y$ slab and the refractive index steps achieved by controlled oxidation or with photo-refractive materials. The device is rotationally symmetric around the propagation axis $z$.}
\label{fig.convertorb}
\end{figure}

The outcome of the calculation and optimization process described earlier is depicted in the schematized Figure \ref{fig.convertor} and the fine resolution ($d=1 nm$) plot of the refractive index in Figure \ref{fig.convertorb}. The refractive index map is symmetric around the propagation axis $z$, and is comprised of a smooth variation of the refractive index between $1$ and $\sim 2.57$. The minimum feature size, $d = d^z = d^y$, is $1nm$, and has the final dimensions $200\mu m$ length by $20\mu m$ wide. Notice that the minimum feature size defines the smoothness of the refractive index map.\\

\begin{figure}[b]
\centering
\subfloat[]{\label{fig.field1a}\includegraphics[width=0.5\textwidth]{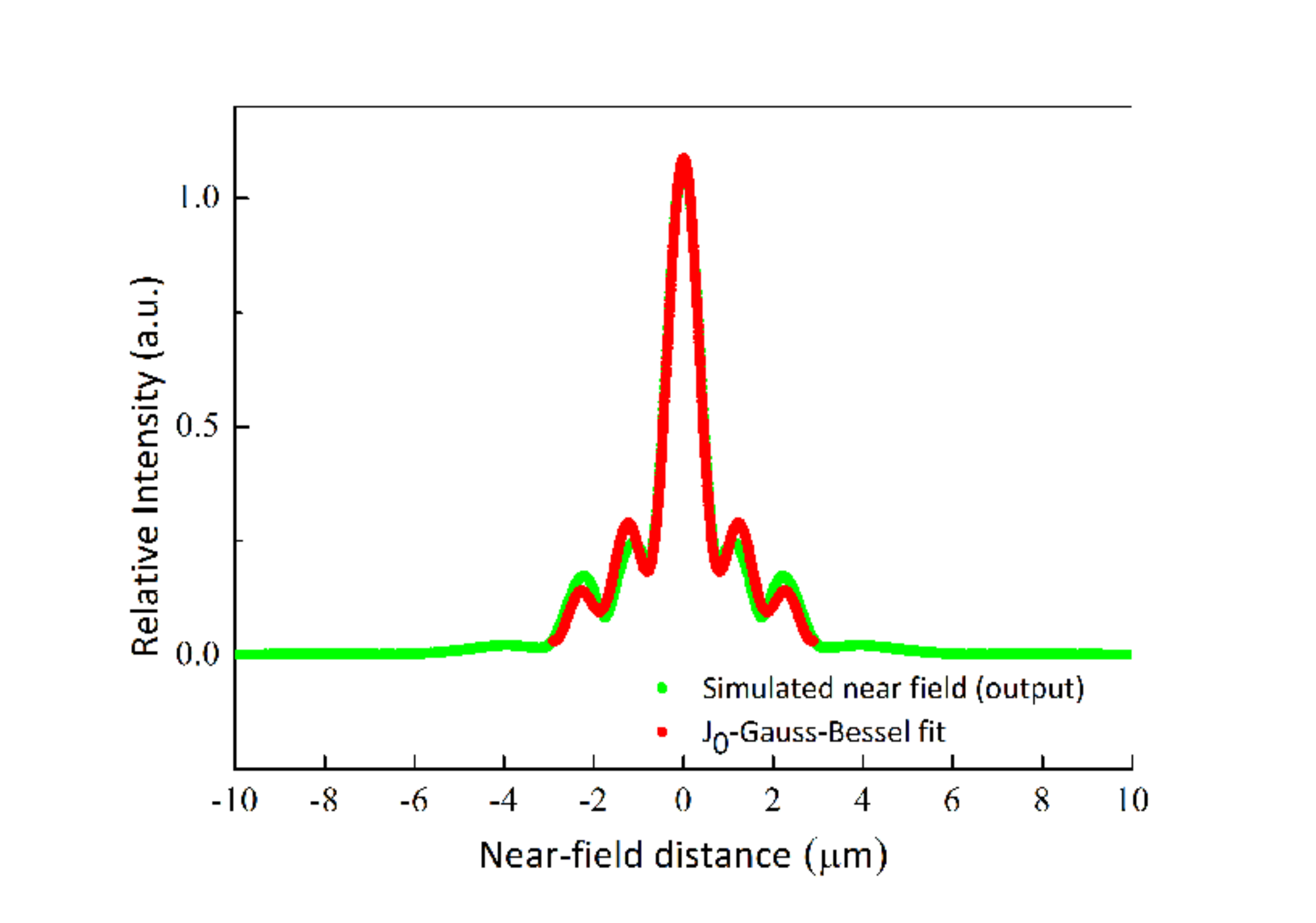}}
\subfloat[]{\label{fig.field1b}\includegraphics[width=0.5\textwidth]{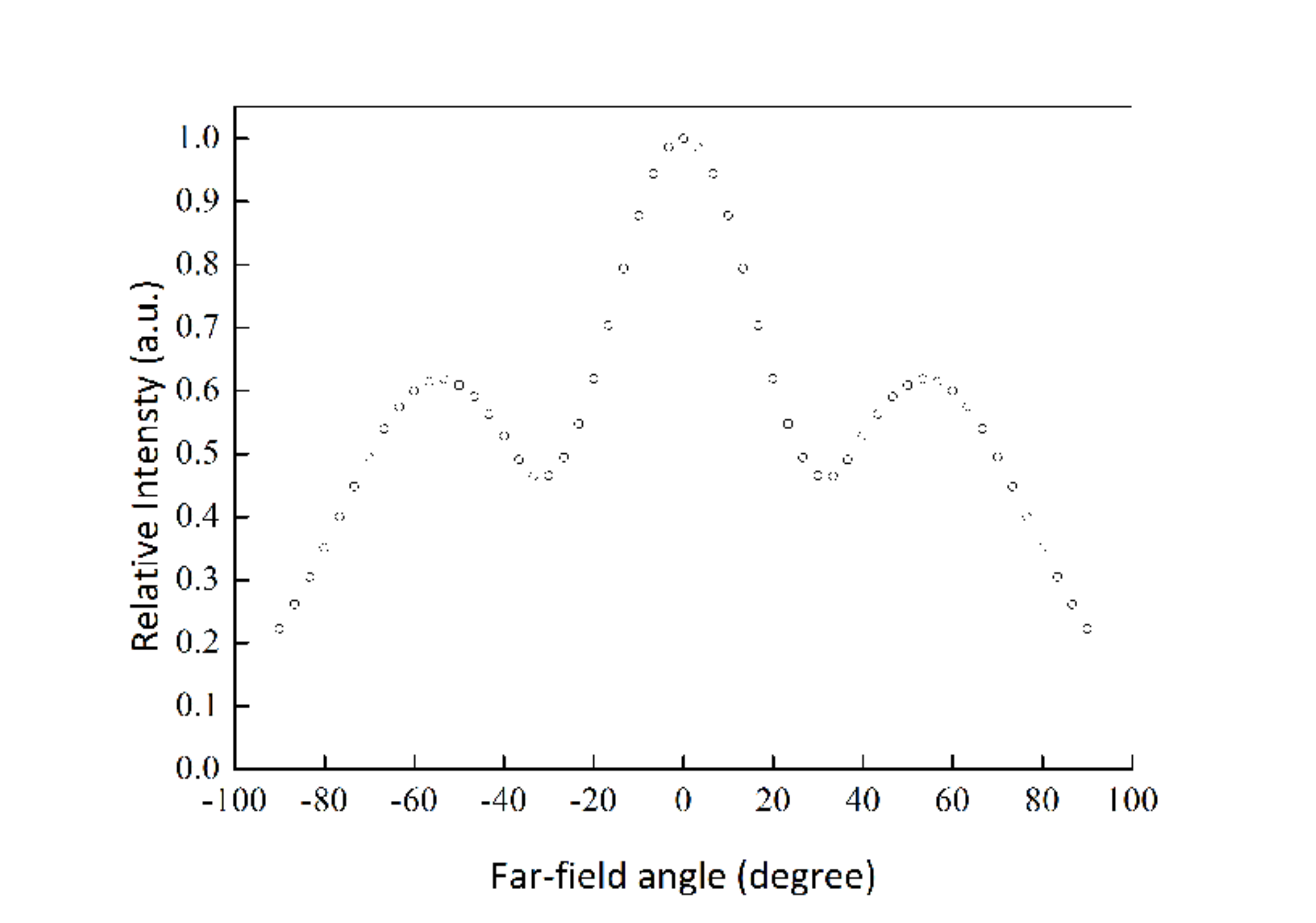}}\\
\subfloat[]{\label{fig.field1c}\includegraphics[width=0.5\textwidth]{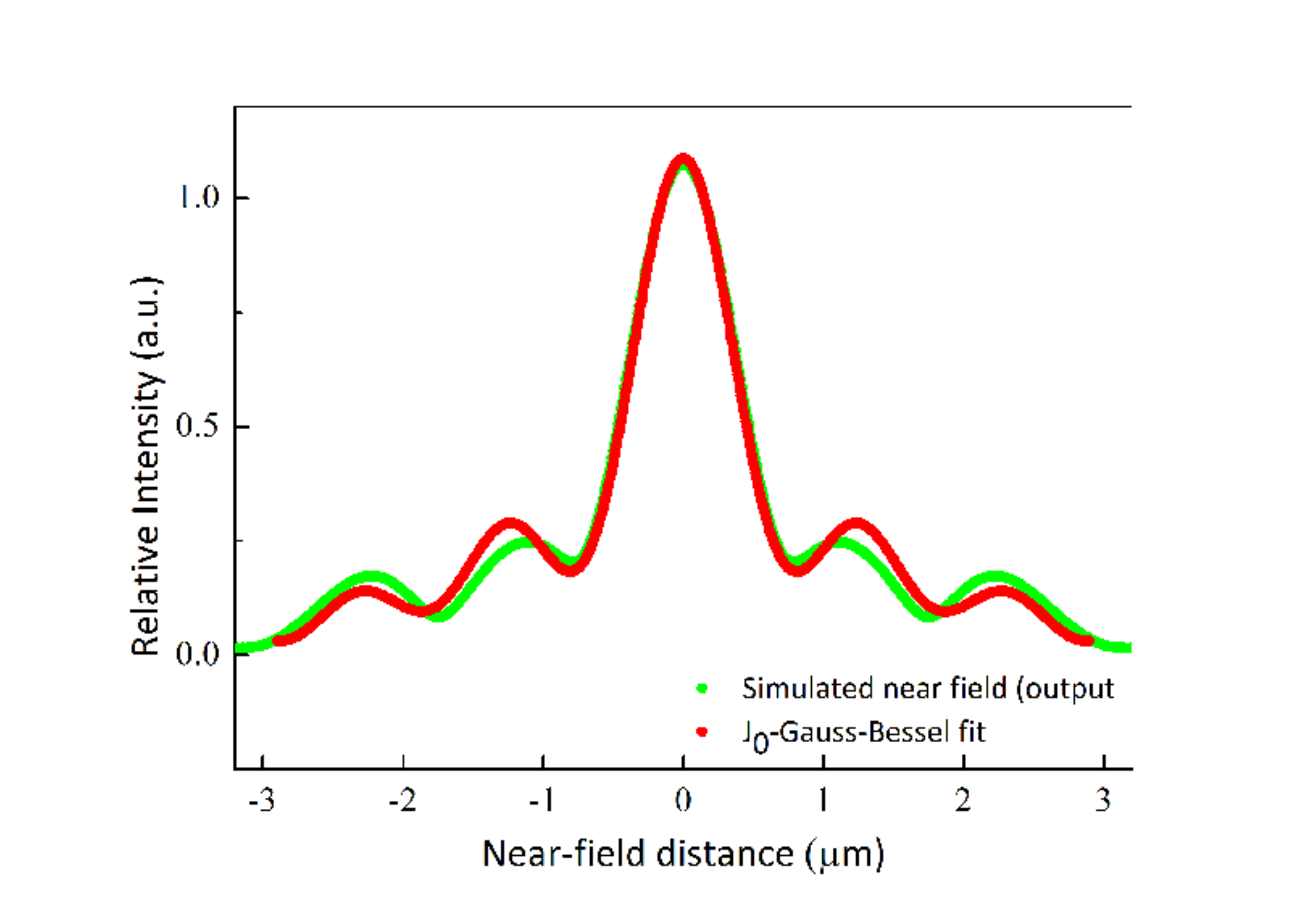}}
\caption{Transverse electric field intensity of the converted beam: (a) simulated near field computed at $5\mu m$ from the output of the device (green), and fit, $95\%$ confidence, to  $J_0$-\bgj (red). Its far field profile (b), and (c) zoom to principal peak.}\label{fig.trans1}
\end{figure}

The initial impinging Gauss beam in region $B_i$  propagates through the device in region $A$, and it is transformed into a \be-\bgj beam with a $95\%$ efficiency. Figure \ref{fig.field1a} shows the results for the field at $5 \mu m$ from the exit of the device, with the corresponding far field depicted in Figure \ref{fig.field1b}. As described earlier, this is the optimized value considering the current state-of-the-art nano-fabrication techniques and facilities available to realize the device.\\

In Figure \ref{fig.trans1}, it is observed that both near and far fields follow a \be-\bgj profile with $98\%$ accuracy. The converted beam divergence is less than $1\%$ after propagating $50 \mu m$ and less than $5\%$ after traveling $1 mm$. The energy density at the principal peak is $95\%$ of the converted beam's total energy, which makes it ideal for OCT applications and free space light-based communication. Losses due to material absorption, scattering and back reflections are $10\%$ of the incident field, or $−0.457dB$, which can be significantly reduced by using antireflection coatings at the input and output surfaces.\\

\begin{figure}[b]
\centering
\subfloat[]{\label{fig.field2a}\includegraphics[width=0.5\textwidth]{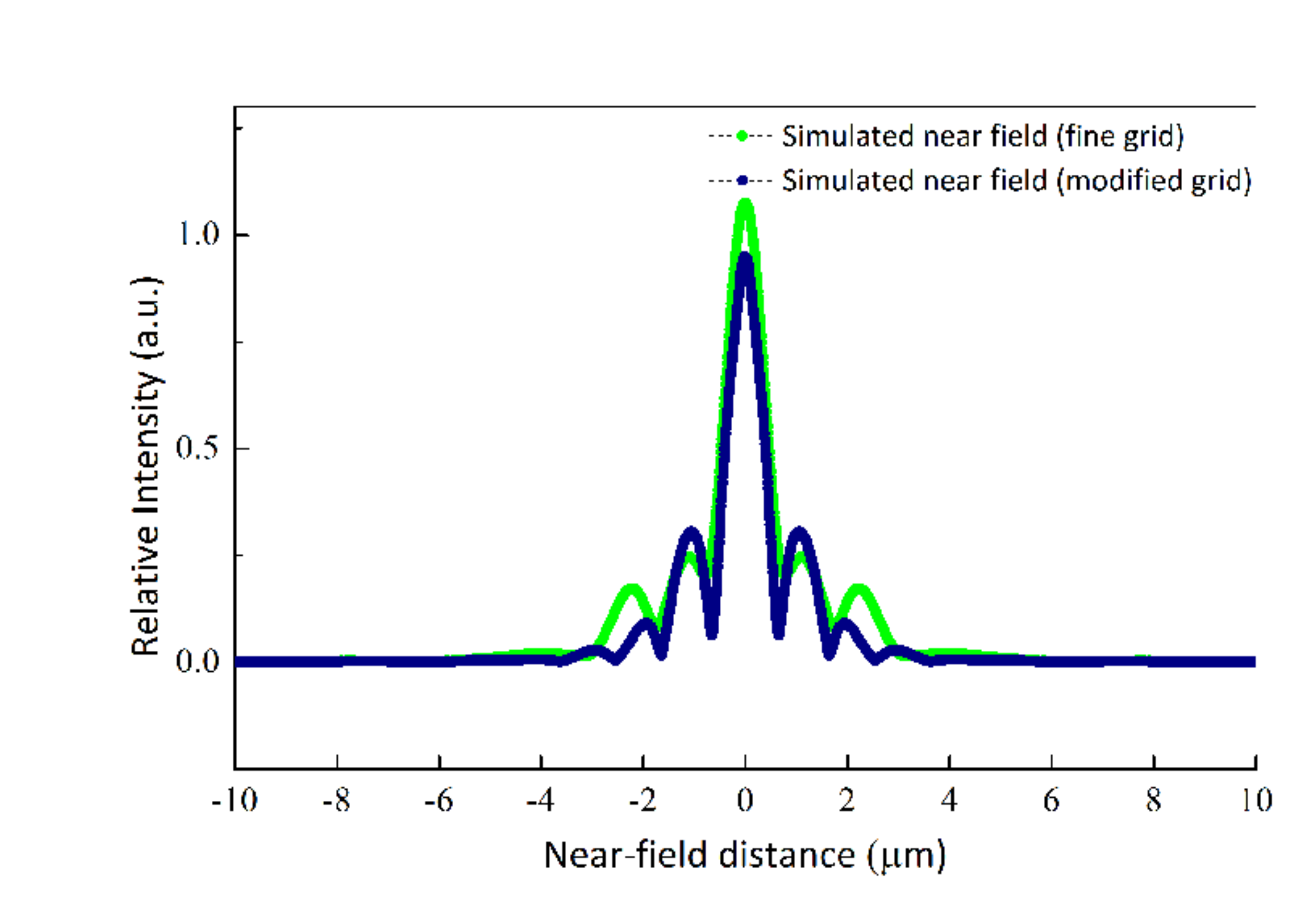}}
\subfloat[]{\label{fig.field2b}\includegraphics[width=0.5\textwidth]{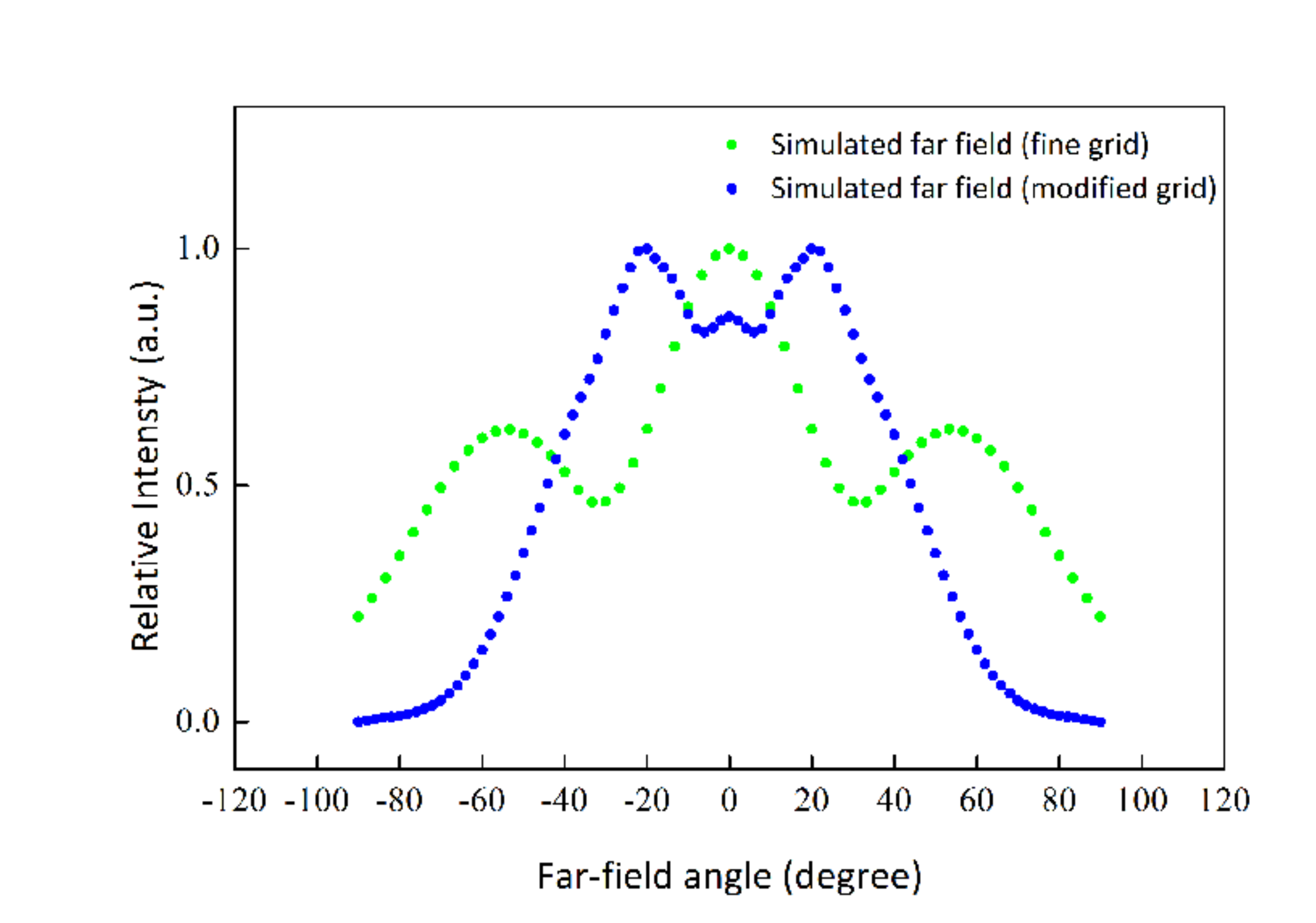}}\\
\subfloat[]{\label{fig.field2c}\includegraphics[width=0.5\textwidth]{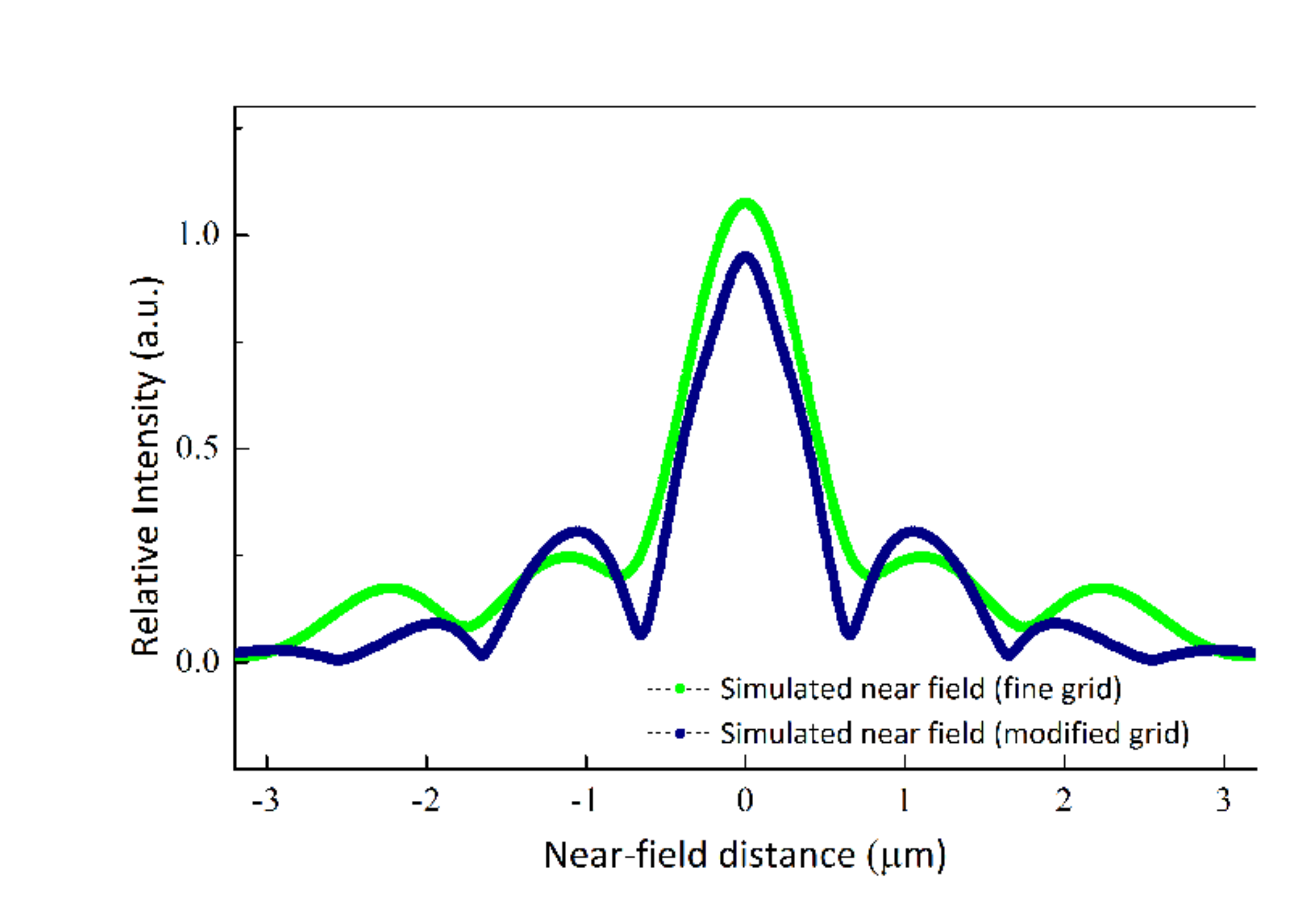}}
\caption{Transverse electric field intensity profile for the converted beam for two different grid sizes: (a) shows the optimal grid size of $1nm$ (green) and modified grid size of $10nm$ (blue), both computed at the  near fields, $5\mu m$ from the output of the device. The far fields for both output beams (b), for optimal grid size (green) and modified grid (blue); and (c) magnification of the near field around the principal peak. \label{fig.trans2}}
\end{figure}

Within the known limitations in fabrication processes, it is noted that not all values of the index of refraction can be achieved. A theoretical study is required to define the tolerance to minimum feature size and its repercussions in the beam profile. We study this case by further modifying the grid size, \ie the minimum feature size, $d = d^z = d^y$ from $1nm$ to $10nm$. The results are shown in Figure \ref{fig.trans2}.\\

As it is evident from Figure \ref{fig.field2a}, modifying the refractive index grid size results in a beam that also resembles a \be-\bgj beam with a $95\%$ accuracy, with the sole difference that this beam as it propagates will evolve into a non-\bgj beam. This can be drawn from the far field, plotted in Figure \ref{fig.field2b}. Observe that at infinity the field no longer corresponds to the far field of a \be-\bgj beam. Nevertheless, a coarse mapping will produce a highly focused beam, where $95\%$ of the energy is localized at the first peak, and the beam diffraction over $1mm$ propagation is less than $5\%$.\\

\begin{figure}[b]
\centering
\subfloat[]{\label{fig.field3a}\includegraphics[width=0.5\textwidth]{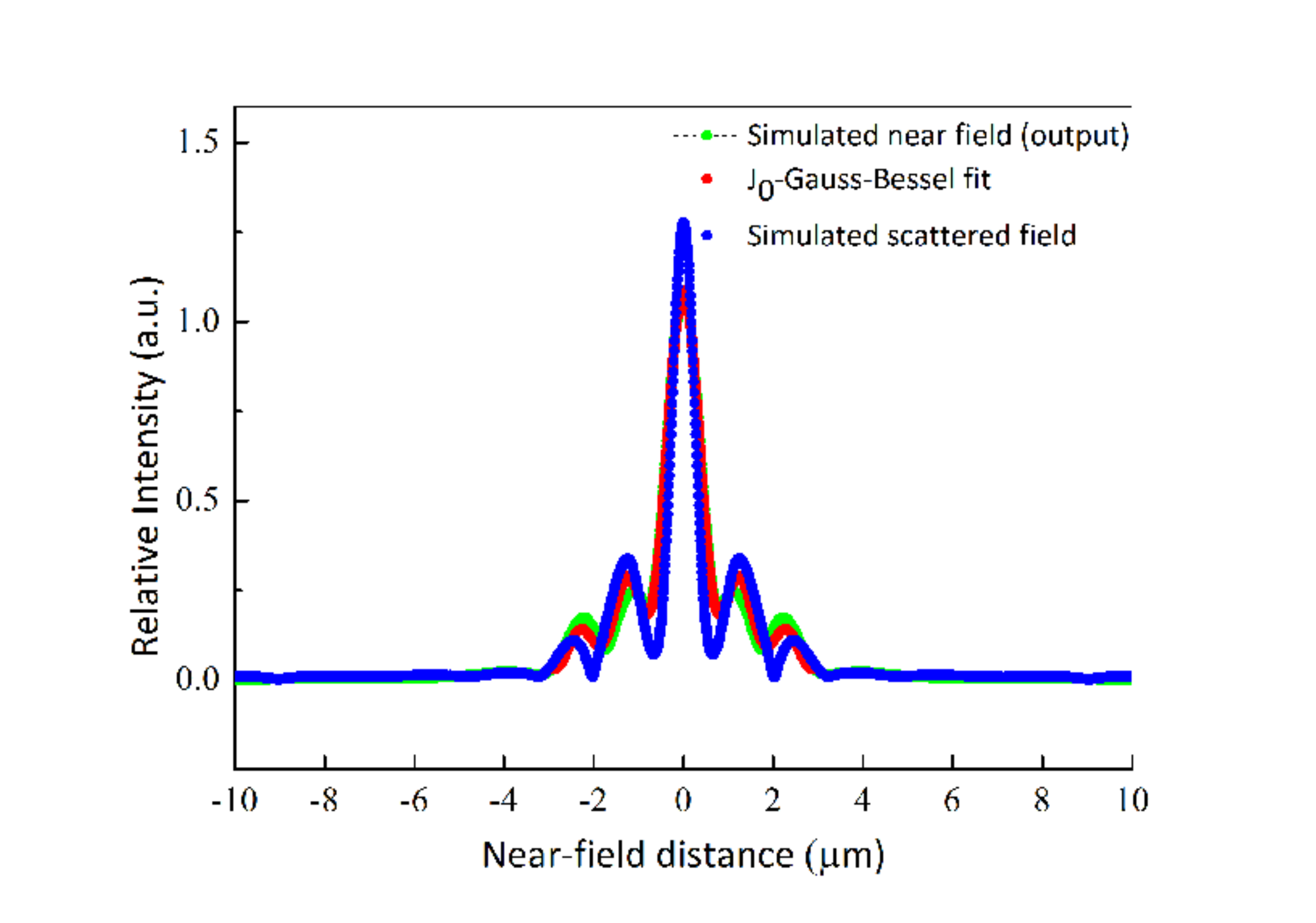}}
\subfloat[]{\label{fig.field3b}\includegraphics[width=0.5\textwidth]{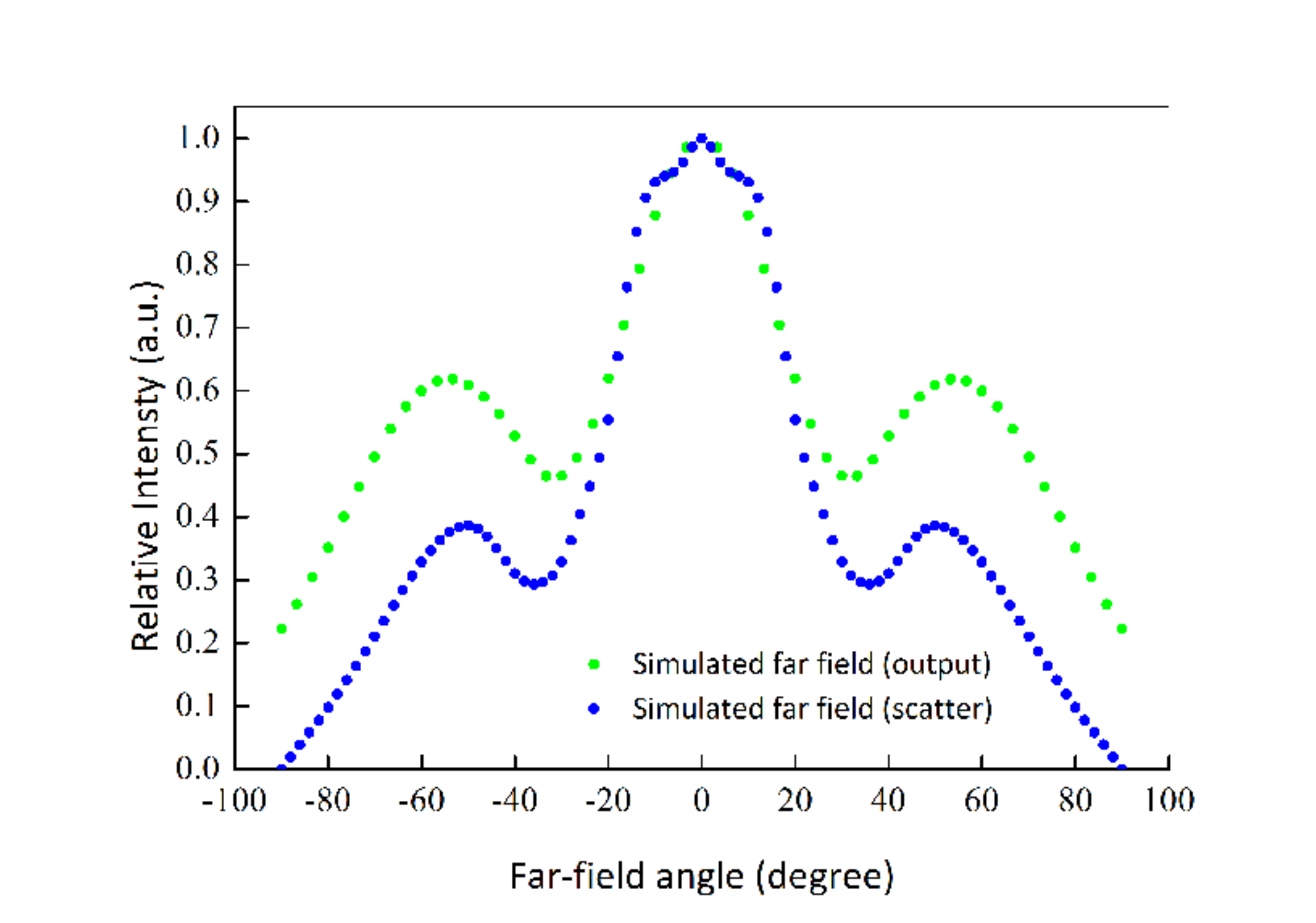}}\\
\subfloat[]{\label{fig.field3c}\includegraphics[width=0.5\textwidth]{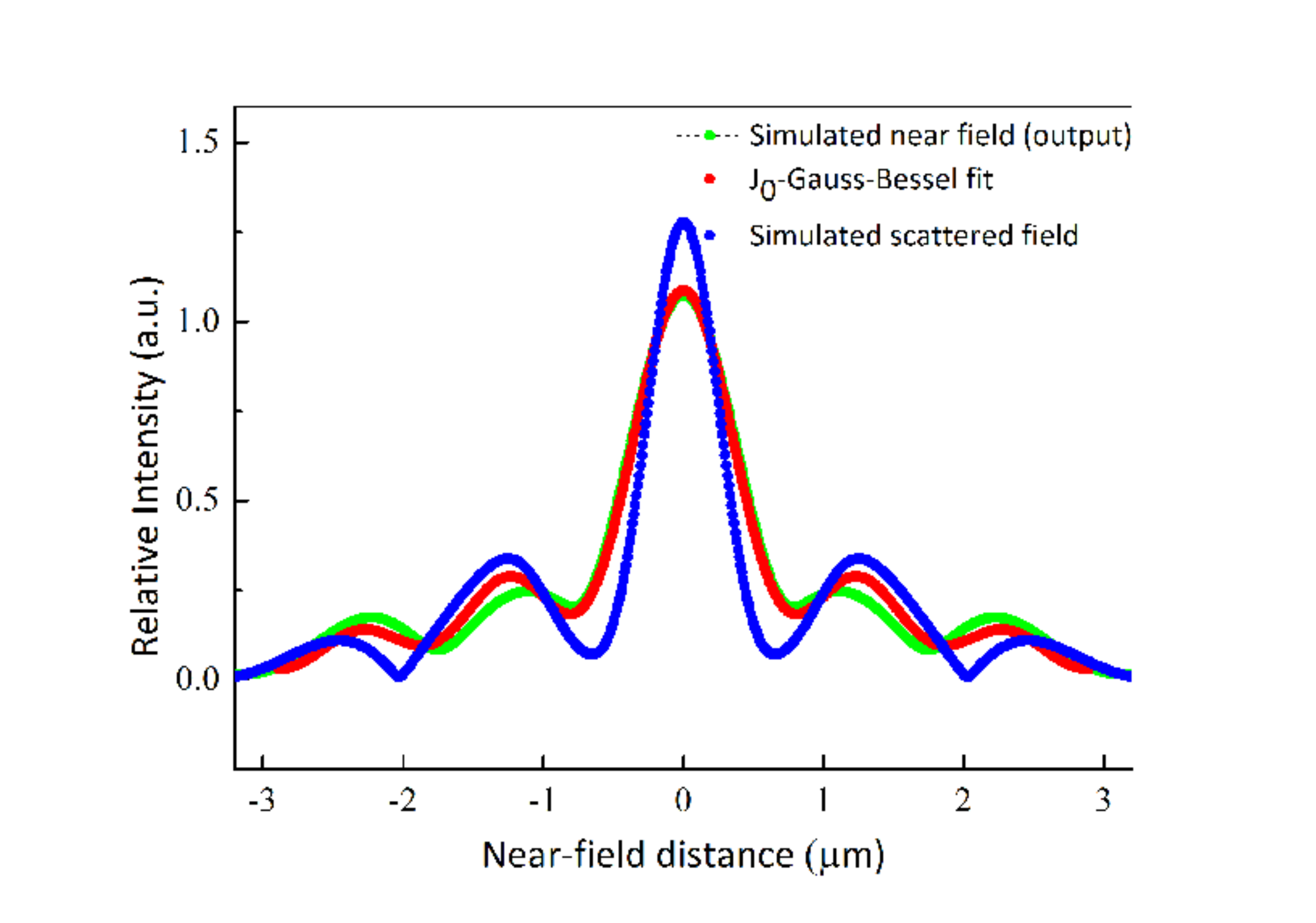}}
\caption{Transverse electric field intensity profile for: (a) the converted beam (green), its fit (red), and the scattered field profile (blue),  computed at $50 \mu m$ from the scattering region. The far field for the undisturbed  and perturbed beams are shown in (b); and (c) magnification of the fields around the principal peak. \label{fig.trans3}}
\end{figure}

\bgj beams, alike Bessel beams, exhibit self regeneration when their path is perturbed by a non- transparent scatterers. To examine this self-healing property of the outbound  \be-\bgj beam, we insert a random index medium on its propagation path.The results are shown in Figure \ref{fig.trans3}.\\

The generated \be-\bgj beam is made to impinge on a random index media, which consists of a circle-like region of $4 \mu m$, set at $50 \mu m$ from the output of the device, and refractive index of $n = 2.0$. The scatterer partially obstructs the propagation of the beam. The scattered beam profile is computed at $50\mu m$ from the alien obstruction (see Figure \ref{fig.field3a} and its magnified version in Figure \ref{fig.field3c}), as well as  the far field (see Figure \ref{fig.field3b}). As can be seen from the results in Figure \ref{fig.trans3}, the beam described earlier (see Figure \ref{fig.trans1}) exhibits self-regeneration. The divergence, computed by comparing the width of the principal peak, from the scattered field to the original transformed beam is $0.3\%$. Since the beam is not a full Bessel beam, some light is scattered by the object, as expected, giving rise to a loss of $19\%$ or $−0.915dB$.\\

{\cte The 2D mapping for beam conversion heretofore described while challenging to fabricate, could be built using current nano-manufacturing techniques. A prototype, \emph{verbi gratia}, could be obtain by controlling the degree of oxidation in $Si/Si_{1−y} O_y /Si_{1−x−y}Ge_x C_y$ \cite{32,33} systems, or by horizontally stacking nano-layers of photo-refractive materials such as chalcogenide glasses which have been shown to have a wide range of refractive index \cite{34,35}. The advantages of these materials are that they are  relatively well-developed, and the material systems are compatible with  PICs and PLCs}.\\

\section{Conclusions}

We have, to the best of our knowledge, theoretically demonstrated the viability of the first PIC and PLC compatible device utilizing a heterogeneous refractive index map, to achieve the transformation of a Gauss beam into a \be-\bgj beam. The computed device has a loss of $−0.457dB$, and high energy focus, where $95\%$ of the output beam energy is concentrated at the principal peak of the \be-\bgj beam profile. The beam has a divergence of $\leq 5\%$ over a propagation distance of $1 mm$, and exhibits self-healing when partially obstructed by an opaque object. The resulting beam has a \be-\bgj profile for both near- and far-fields. It is note that our design is based on current manufacturing techniques, such as controlled oxidation of $Si/Si_{1−y} O_y /Si_{1−x−y}Ge_x C_y$ or by stacking different layers of chalcogenide glasses. An integrated Gauss to \be-\bgj beam convertor could enable on-chip applications which take advantage of the beam’s self-healing and non-diffractive properties, such as micro optical tweezers, traps and couplers, photonic integrated circuits for telecommunications and computerized micro tomography and microscopy.\\

\bibliographystyle{osajnl}

\end{document}